\documentclass[11pt]{article}

\usepackage[margin=1in]{geometry}
\usepackage{setspace}
\usepackage[ruled,vlined]{algorithm2e}
\usepackage[T1]{fontenc}
\usepackage[utf8]{inputenc} 
\usepackage{lmodern}

\usepackage{amsmath,amssymb,amsfonts}
\usepackage{bm}

\usepackage{graphicx}
\usepackage[skip=8pt]{caption}
\usepackage{float}
\usepackage{booktabs}
\usepackage{multirow}

\usepackage{natbib}

\usepackage{hyperref}

\usepackage[acronym]{glossaries}

\newacronym{ssm}{SSM}{State Space Model}

\newacronym{ibis}{IBIS}{Iterated Batch Importance Sampling}
\newacronym{smc}{SMC}{Sequential Monte Carlo}
\newacronym{smc2}{SMC\textsuperscript{2}}{Sequential Monte Carlo Squared}

\newacronym{mcmc}{MCMC}{Markov Chain Monte Carlo}
\newacronym{hmc}{HMC}{Hamiltonian Monte Carlo}
\newacronym{nuts}{NUTS}{No U-Turn Sampling}
\newacronym{pmh}{PMH}{Particle Metropolis Hastings}
\newacronym{pgibbs}{PGIBBS}{Particle Gibbs}
\newacronym{pgas}{PGAS}{Particle Gibbs with ancestors sampling}
\newacronym{pmcmc}{PMCMC}{Particle MCMC}

\newacronym{pf}{PF}{Particle Filter}
\newacronym{cpf}{CPF}{Conditional Particle Filter}
\newacronym{bpf}{BPF}{Bootstrap Particle Filter}
\newacronym{apf}{APF}{Auxiliary Particle Filter}

\newacronym{hmm}{HMM}{Hidden Markov Model}
\newacronym{hsmm}{HSMM}{Hidden Semi-Markov Model}
\newacronym{edhmm}{EDHMM}{Explicit-duration Hidden Markov Model}
\newacronym{arhsmm}{ARHSMM}{Autoregressive Hidden Semi-Markov Model}

\newacronym{hsmm-egarch}{HSMM-eGARCH}{Hidden Semi-Markov Model with eGarch volatility dynamics}

\newacronym{svm}{SVM}{Stochastic Volatility Model}
\newacronym{mjm}{MJM}{Markov Jump Model}
\newacronym{garch}{GARCH}{Generalized Autoregressive Conditional Heteroskedasticity}
\newacronym{egarch}{eGARCH}{Exponential Generalized Autoregressive Conditional Heteroskedasticity}

\newacronym{crps}{CRPS}{Continuous Ranked Probability Score}
\newacronym{bmi}{BMI}{Basic Marginal Likelihood Identity}
\newacronym{is}{IS}{Importance Sampling}
\newacronym{sis}{SIS}{Sequential Importance Sampling}
\newacronym{sir}{SIR}{Sequential Importance Resampling}

\newacronym{seir}{SEIR}{Susceptible-Exposed-Infected-Recovered}
\newacronym{seeiir}{SEEIIR}{Susceptible-Exposed-Exposed-Infected-Infected-Recovered}
\newacronym{ode}{ODE}{Ordinary Differential Equation}
\newacronym{rt}{$R_t$}{Effective Reproductive Number}
\newacronym{hsmmem}{HSMM-EM}{HSMM-driven epidemic Model}
\newacronym{ifr}{ifr}{Infection Fatality Ratio}
\newacronym{gbm}{GBM}{Geometric Brownian Motion}

\newacronym{ci}{CI}{Credible Interval}
\newacronym{dic}{DIC}{Deviance Information Criterion}
\newacronym{waic}{WAIC}{Watanabe–Akaike Information Criterion}
\newacronym{lppd}{LPPD}{Log Pointwise Predictive Density}
\newacronym{dag}{DAG}{Directed Acyclic Graph}
\newacronym{pl}{PL}{Predictive Likelihood}
\newacronym{bf}{BF}{Bayes Factor}
\newacronym{clpbf}{CLPBF}{Cumulative Log Predictive Bayes Factor}

\usepackage{float}

\begin{document}


\title{Semi-Markov Models with Particle-Based Bayesian Inference for Epidemics}


\author{
Patrick Aschermayr$^{1}$, 
Konstantinos Kalogeropoulos$^{1}$, 
Nikolaos Demiris$^{2}$\\[6pt]
\small $^1$London School of Economics and Political Science\\
\small $^2$Athens University of Economics and Business\\[4pt]
\small \texttt{k.kalogeropoulos@lse.ac.uk}
}

\date{}

\maketitle

\begin{abstract}
The COVID-19 pandemic has been characterised by multiple waves of transmission driven by interventions and emerging variants, challenging epidemic models that assume gradually evolving transmission dynamics. We propose a class of state-space models in which the transmission rate evolves through persistent regimes of random duration, governed by a semi-Markov process. This formulation yields an interpretable representation of sustained transmission phases and retains a parsimonious parameterisation. Particle-based Bayesian methods are well established for standard state-space models, but their use in semi-Markov settings has received comparatively limited attention. In epidemic applications, inference is further complicated by differential equation-driven latent dynamics and observation models defined through functionals of the latent process. We develop an inferential framework that accommodates these features, combining particle-based state updates with gradient-based parameter updates and enabling batch and sequential inference via particle and sequential Monte Carlo. We apply the proposed methodology to COVID-19 data from the United Kingdom and show that combining reported cases and deaths leads to more precise and stable inference compared to using deaths alone. These results illustrate the practical value of semi-Markov transmission models for epidemic analysis under complex observation schemes.
\end{abstract}



\section{Introduction} \label{sec:cov_Introduction}

Since the emergence of the SARS-CoV-2 virus in late 2019, policy-makers have faced substantial challenges in designing timely intervention strategies to control COVID-19 outbreaks. Due to the high reproduction rate of the virus, it is crucial to monitor the current number of infections in order to prevent uncontrolled spread and its associated societal and economic consequences. \cite{Petrosillo20} and \cite{Liu20} provide estimates for various coronavirus transmission rates, placing even early SARS-CoV-2 variants among the most infectious.

The standard framework for modelling such phenomena is given by compartmental models, which partition the population into different groups. For example, the well-known \acrfull{seir} model, see, e.g., \citep{Diekmann12}, divides the population into susceptible, exposed, infectious, and recovered individuals. A key advantage of this class of models is their interpretability and their suitability for studying the impact of interventions. An important quantity derived from \acrshort{seir}-type models is the \acrfull{rt}, representing the expected number of secondary infections generated by an infected individual over the course of their infectious period. In the standard setting, this quantity is driven by the transmission rate $\beta$, which is typically assumed constant over time. However, for SARS-CoV-2 transmission it has become apparent that $\beta$ varies over time, reflecting behavioural changes and policy interventions; see, e.g., \cite{Flaxman20}. Consequently, standard \acrshort{seir} models with constant transmission rates are not well suited for accurate inference and prediction.

A common approach to address this limitation is to allow $\beta$ to vary over time. Existing approaches broadly fall into two categories: models based on discrete change points, see, e.g., \cite{Barbounakis}, and models based on continuous latent processes, such as \acrfull{gbm}, see, e.g., \cite{Kalogeropoulos13,xu18}. Change-point models are highly interpretable but typically describe transmission dynamics through abrupt shifts, rather than allowing persistent within-regime evolution and temporal dependence. Continuous latent processes provide smoother dynamics, but typically lead to less direct interpretation in terms of distinct transmission regimes such as suppression, transitional, or high-transmission phases.

We propose a class of \acrshort{seir}-type models that combines the advantages of both approaches by modelling the transmission rate as piecewise constant and driven by a latent \acrlong{hsmm}. This provides an interpretable representation of distinct transmission regimes while explicitly modelling their duration and transition dynamics. In contrast to standard hidden Markov models, the semi-Markov formulation allows non-geometric dwell-time distributions and therefore accommodates persistent epidemic phases more realistically. The state-space formulation further allows the computation of transition probabilities between regimes, offering additional insight into the evolution of the epidemic. Incorporating such a latent structure substantially increases the computational complexity of the model. To address this, we develop a tailored \acrfull{pf} for latent state inference and combine it with a \acrlong{pmcmc} kernel for batch estimation of model parameters. In addition, we employ a \acrfull{smc2} algorithm for sequential inference and prediction.

Our main contributions are (i) the formulation of a flexible and interpretable state-space model for epidemic dynamics that identifies persistent latent transmission regimes, (ii) the development of efficient inference procedures for such processes, and (iii) the assessment of the added predictive ability of distinct data sources. The latter is evaluated by assessing the predictive performance on COVID-19 infections and fatalities in the UK, using these results as a basis for model comparison and for identifying the number of latent regimes supported by the data. We show that combining reported infections and fatalities improves predictive performance compared to models based solely on fatalities. The remainder of the paper is organised as follows. Section \ref{sec:cov_Model} introduces the proposed model. Section \ref{sec:cov_BayesianInference} presents the inference methodology. Section \ref{sec:cov_Application} applies it to COVID-19 data from the UK, and Section \ref{sec:cov_Conclusion} concludes with a discussion.

\section{Model} \label{sec:cov_Model}
\subsection{Epidemic model specification}   

We link model-implied infections and deaths to the corresponding reported data. Throughout this paper, we will use a superscript r for the reported data, and a superscript i for the model-implied data. The model-implied cases are obtained from the following system of \acrlong{ode}s, which is an extension of a standard \acrshort{seir} Model:
\begin{equation}
\begin{split}
\frac{\mathrm{d}S_t}{\mathrm{d}t} & = - \beta_t S_t \frac{\left(I_{1,t}+I_{2,t}\right)}{N} - \rho \nu_{t-U},\\
\frac{\mathrm{d}E_{1,t}}{\mathrm{d}t} & =  \beta_t S_t \frac{\left(I_{1,t}+I_{2,t}\right)}{N} - \epsilon E_{1,t},\\
\frac{\mathrm{d}E_{2,t}}{\mathrm{d}t} & = \epsilon E_{1,t} - \epsilon E_{2,t},\;\;
\frac{\mathrm{d}I_{1,t}}{\mathrm{d}t} = \epsilon E_{2,t} - \gamma I_{1,t},\\
\frac{\mathrm{d}I_{2,t}}{\mathrm{d}t} & = \gamma I_{1,t} - \gamma I_{2,t},\;\;
\frac{\mathrm{d}R_{t}}{\mathrm{d}t}  =  \gamma I_{2,t} +\rho \nu_{t-U},
\label{eq:cov_Chp1_ode_SEEIIR}
\end{split}
\end{equation}
The additional compartments for the Exposed and Infectious states, $E_2$ and $I_2$, result in more realistic, Erlang-distributed, incubation and infectious time periods, e.g. \cite{Wearing05}. We denote by $\nu_t$ the number of people who received their first COVID-19 vaccination at time $t$. Vaccine effects are incorporated through a delay parameter $U$ and an efficacy parameter $\rho$.

In addition to the model parameters stated in Equations \eqref{eq:cov_Chp1_ode_SEEIIR}, inference requires estimating a discrete latent state trajectory of length $T$, where $T$ is the number of observations. The latent state trajectory $s_{1:T}$ determines the time-varying $\beta_t$.  The model-implied deaths $d^i$ are defined based on the model-implied cases $c^i$, as
\begin{equation}
    d^i_{t} = \text{ifr}_t * \sum_{\tau=max(1, t-28)}^{t-1} c^i_{\tau} f_{t-\tau},
\label{eq:cov_Model_deathsimplied}
\end{equation}
where $d^i_0$ is set to 0; see \cite{Flaxman20, Chatzilena22}. Here $f$ is the distribution of time from infection to death and is based on the estimates of \cite{verity20} while \acrfull{ifr} denotes the probability of death for an infected individual and is based on \cite{levin20} as adjusted by \cite{Chatzilena22}. We assume fatalities occur within $28$ days of infection, consistent with reporting practices in the UK, see \cite{Coviddata22}. The observation model for the reported cases is defined as
\begin{equation}
    c^r_{t} \sim \operatorname{Negative \hspace{0.1cm} Binomial}_{Alternative} \left(c^*_t,c^*_t + \frac{c^{*2}_{t}}{\phi_c}\right),
\label{eq:cov_Model_casesreported}
\end{equation}
where $c^*_t = c^i_t * ur_t$ are the model implied cases adjusted by an under-reporting score that is provided externally and varies across time. The $\operatorname{ Negative \hspace{0.1cm} Binomial}_{Alternative}(\mu, \phi)$  parameterisation of the Negative Binomial distribution has $\mathop{ \mathbb{E}}(Y) = \mu$ and $Var(Y)= \mu + \frac{\mu^2}{\phi}$. 
Similarly, the observation model for the reported fatalities is
\begin{equation}
    d^r_t \sim \operatorname{Negative \hspace{0.1cm} Binomial}_{Alternative} \left(d^i_t,d^i_t + \frac{{d^i}^{2}_{t}}{\phi_d}\right).
\label{eq:cov_Model_deathsreported}
\end{equation}
The $c^i$ values are derived from the ODE system in Equation \eqref{eq:cov_Chp1_ode_SEEIIR} and represent model-implied incidence. To solve these equations, a time-indexed sequence of piecewise constant transmission rates $\beta_t$ is required. We obtain these by linking each $\beta$ to a latent variable that follows a \acrshort{hsmm}.

\begin{figure}[htp]
\centering
    \includegraphics[width=0.93\textwidth]{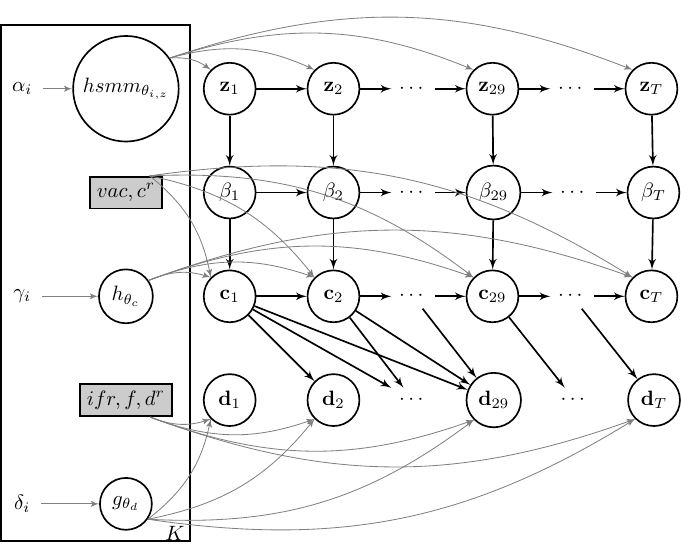}
    \caption{Graphical representation of the \acrlong{hsmm} with parameters $\theta$ and hyper-parameters $\{\alpha,\gamma,\delta\}$. The latent process $z_t$ follows a \acrlong{hsmm} and determines the transmission rate $\beta_t$. Unshaded nodes $c_t$ and $d_t$ denote unobserved true cases and deaths. The quantity $\theta_{i,z}$ is the parameter associated with regime $i$ and state $z$. Functions $h$ with parameters $\theta_c$ govern model-implied cases, while $g$ with parameters $\theta_d$ governs model-implied deaths. Variables $vac$, $c_r$, and $d_r$ denote reported vaccinations, cases, and deaths; \acrshort{ifr} is the infection-fatality ratio, and $f$ the infection-to-death delay distribution.}

\label{fig:cov_DAG}
\end{figure}

The \acrshort{hsmm} formulation is also known as the \acrfull{edhmm}, i.e. a \acrshort{hsmm} that explicitly defines the distribution of the duration remaining in a state. Transitions are allowed only at the end of each state, resulting in the following definition. A hidden semi-Markov model is a bivariate stochastic process $ \{ e_t, z_t \}_{t = 1,2,\ldots} $, where $z_t = \{ s_t, d_t \}$ is an unobserved semi-Markov chain and, conditional on $z_t$, $e_t$ is an observed sequence of independent random variables. The model is fully specified by the transition distribution $f_\theta( s_t \mid s_{t-1}, d_{t-1} )$:
	\begin{equation} 
    s_t \sim  \begin{cases}
    \delta( s_{t}, s_{t-1}) &\text{ $d_{t-1} > 0$ }\\
    f_\theta( s_t \mid s_{t-1}, d_{t-1} ) &\text{ $d_{t-1} = 0$ },
    \end{cases}
    \label{eq:cov_Chp3_EDHMM_transition}
    \end{equation}
    the duration distribution $h_{\theta}$ of $d_t$
    \begin{equation} 
    d_t \sim \begin{cases} 
    \delta( d_{t}, d_{t-1} - 1) &\text{ $d_{t-1} > 0$ }\\
    h_\theta( d_t \mid s_{t}, d_{t-1}) &\text{ $d_{t-1} = 0$ },
    \end{cases}
    \label{eq:cov_Chp3_EDHMM_duration}
    \end{equation}
the initial density $\pi_{\theta}$ of $z_t$, and the observation distribution $g_{\theta}$, $e_t \sim g_{\theta}(e_t \mid s_t)$,
    \begin{equation} 
    e_t \sim g_\theta(e_t \mid s_{t}). 
    \label{eq:cov_Chp3_EDHMM_observation}
    \end{equation}
    where $\delta(a,b)$ denotes the Kronecker delta, i.e. equals $1$ if $ a = b$ and $0$ otherwise.
We model state durations using a Negative Binomial distribution for additional flexibility. A review of \acrshort{hsmm}s can be found in \cite{yu10, Yu16}; see also \cite{Bulla06, Lindsten13, Papaspiliopoulos20, corenflos21} for additional applications. 
Figure \ref{fig:cov_DAG} depicts the \acrfull{dag} of the \acrshort{hsmmem}.

\subsection{From epidemic dynamics to a semi-Markov state-space representation}

Epidemic dynamics during COVID-19 are often characterised by abrupt changes driven by interventions or behavioural shifts, suggesting that the transmission rate $\beta_t$ evolves through a sequence of regimes rather than as a smooth process. We model this by letting $\beta_t$ be piecewise constant, with values determined by a latent process following a hidden semi-Markov model. This allows for explicit modelling through flexible duration distributions and results in interpretable distinct epidemic phases. Let $O_t$ denote the epidemic state at time $t$ and $z_t=(s_t,d_t)$ the semi-Markov regime and its remaining duration. Then $\beta_t$ is determined by $s_t$, and the latent state is given by $X_t = (O_t, z_t)$. The state evolution can be written as
\[
z_{t+1} \sim p(z_{t+1} \mid z_t), \qquad 
O_{t+1} = \Phi(O_t, \beta_{t+1}, \theta),
\]
where $\Phi(\cdot)$ denotes the one-step solution of the ODE system in Equation \eqref{eq:cov_Chp1_ode_SEEIIR}. Conditional on the latent state, reported cases depend only on the current model-implied incidence via
\[
c^r_t \sim p_\theta(c^r_t \mid O_t).
\]

While reported cases depend primarily on the current latent state, reported deaths depend on past infections through a convolution over a window of length $x=28$. To accommodate this, we augment the latent state as
\[
\tilde X_t = \bigl(O_t,\; z_t,\; c^i_{t-x+1:t}\bigr),
\]
where $c^i_{t-x+1:t} = (c^i_{t-x+1},\dots,c^i_t)$.
The lagged infection component evolves via a deterministic sliding-window update,
\[
c^i_{t-x+2:t+1} = (c^i_{t-x+2},\dots,c^i_t,c^i_{t+1}),
\]
where the new incidence term satisfies
\[
c^i_{t+1} = h(O_t, \beta_{t+1}, \theta),
\]
with $h(\cdot)$ denoting the mapping from the ODE solution to model-implied infections. Under this augmented representation, the transition takes the form
\[
\tilde X_{t+1} = \Psi(\tilde X_t, z_{t+1}, \theta),
\]
where $\Psi(\cdot)$ combines the semi-Markov update for $z_{t+1}$, the ODE evolution, and the deterministic shift of the infection history.
With this augmented state, the process is Markov in $\tilde X_t$, even though the observation model depends on past latent quantities. This formulation underpins the inference methods developed in the following section.

\section{Particle-Based Bayesian Inference} \label{sec:cov_BayesianInference}

\subsection{Challenges in posterior inference}

Posterior inference in the proposed model is complicated by the structure of both the latent process and the observation model. In standard hidden Markov or semi-Markov models with discrete state spaces and tractable emission distributions, inference can often be performed by marginalising over the latent states using forward–backward algorithms. However, this approach becomes impractical in the present setting. The latent process combines a semi-Markov regime component with continuous epidemic dynamics governed by an ODE system, while the observation model for deaths depends on a delayed functional of past latent infections. As a result, the effective state space includes both continuous components and a history-dependent structure, leading to a high-dimensional augmented state. 

Given that exact marginalisation of the latent states is not feasible, inference proceeds by sampling from the joint posterior distribution of latent trajectories and model parameters. This leads to a high-dimensional inference problem in which the latent state includes both a discrete semi-Markov regime sequence and continuous epidemic variables evolving over time. Sampling from this joint posterior presents several challenges. First, the semi-Markov regime process is discrete and exhibits strong temporal dependence through its duration structure, which precludes the use of gradient-based methods such as Hamiltonian Monte Carlo for updating the latent states. Second, the latent trajectory is high-dimensional, and no natural low-dimensional blocking structure is available for efficient joint updates, leading to poor mixing for naive local proposals. Third, the augmented state required to accommodate the history-dependent observation model further increases dimensionality and introduces additional dependence across time.

These considerations motivate the use of particle-based methods, which allow for joint inference on latent trajectories and model parameters in settings where exact marginalisation is infeasible. In particular, particle MCMC methods provide a principled framework for sampling from the posterior distribution of high-dimensional latent paths, while sequential Monte Carlo squared enables sequential inference and predictive assessment in a coherent Bayesian framework. In the following sections, we develop a particle-based inference framework for batch and sequential inference, combining particle MCMC methods for joint state and parameter estimation with sequential Monte Carlo squared for online inference and predictive assessment.

\subsection{Particle-based inference} \label{subsec:Parameterestimation}

Inference proceeds by sampling from the joint posterior distribution of the model parameters $\theta$ and the latent state trajectory $s_{1:T}$ given the observed data $e_{1:T}$. We develop an inference framework based on \acrlong{pmcmc}, see \cite{Andrieu10}, tailored to the proposed model structure. The combination of semi-Markov latent dynamics, ODE-driven state evolution, and history-dependent observations requires adaptations of standard particle methods. In particular, the latent state sequence must be inferred jointly with model parameters in a setting where exact marginalisation is not feasible.

In standard PMCMC schemes such as the \acrfull{pmh} kernel, parameter proposals are accepted based on particle filter likelihood estimates, which can lead to poor mixing and make the design of efficient proposal kernels challenging. We instead use the \acrfull{pgibbs} variant, which alternates between sampling the latent trajectory via a \acrfull{cpf} and updating the model parameters via an \acrshort{mcmc} kernel. The conditional particle filter preserves a reference trajectory, ensuring that the correct target distribution is maintained; see Section 1 of the Supplementary Material for more details. A key advantage of this approach is that, conditional on the latent states, the likelihood can be evaluated pointwise, allowing the use of more advanced \acrshort{mcmc} kernels for parameter updates. 

For the parameter updates, we use \acrfull{hmc} in its adaptive \acrfull{nuts} variant \cite{neal10, Gelman14}. The particle filter needs to account for the dependence of deaths on past infections, which introduces a finite memory into the observation model. As a result, the incremental weights depend on a window of recent latent states rather than only the current state. While this increases the computational cost relative to standard \acrshort{ssm}s, the overall complexity remains linear in the time horizon $T$. In our experiments, Particle Gibbs schemes with HMC updates for the static parameters proved substantially more stable than Particle MCMC schemes based on Metropolis--Hastings proposals, particularly in higher-dimensional specifications. The algorithm is outlined below:

\bigskip

\begin{algorithm}[H]
	\SetKwData{Left}{left}\SetKwData{This}{this}\SetKwData{Up}{up}
	\SetKwFunction{Union}{Union}\SetKwFunction{FindCompress}{FindCompress}
	\SetKwInOut{Input}{input}\SetKwInOut{Output}{output}\SetKwInOut{Tuning}{tuning parameter}\SetKwInOut{Function}{function} 
	\Input{Reference trajectory $s_{1:T}$, data $e_{1:T}$, current model parameter $\theta$}
	\Output{$(\theta,s_{1:T}) \sim p(\theta,s_{1:T}\mid e_{1:T})$}
    \Tuning{conditional particle filter $cpf$, HMC/NUTS kernel}
    \Function{Particle Gibbs kernel $K_{pgibbs}(e_{1:T}, s_{1:T}, \theta)$}
    \BlankLine

    Update static parameters using an HMC/NUTS step targeting
    $p(\theta\mid s_{1:T}, e_{1:T})$ to obtain $\theta^\star$.
     
	Run the conditional particle filter $cpf$ with parameter $\theta^\star$ to obtain
	$s^{\star}_{1:T}\sim \hat{p}_{\theta^\star}(s_{1:T}\mid s_{1:T}, e_{1:T})$.
	
	Set $(\theta,s_{1:T}) := (\theta^\star,s^{\star}_{1:T})$.
    
    \KwRet $(\theta,s_{1:T})$.
	\caption{Particle Gibbs with HMC parameter updates}
	\label{alg:PGibbs}
\end{algorithm}

\bigskip

Given that data for our model are typically acquired sequentially, we also employ the \acrfull{smc2} framework, see \cite{chopin2012}. This method combines particle filtering with PMCMC updates across multiple parameter particles, allowing parameter estimates to be updated as new data become available. The approach builds on the \acrlong{smc} framework of \cite{Chopin02, DelMoral06}; see also \cite{Dai22} for a review. Within the rejuvenation step, tuning \acrfull{hmc} separately for each particle would be computationally costly. We therefore estimate a common mass matrix from the current particle population and tune the step size and leapfrog length using a representative particle, reusing these settings across particles. The SMC$^2$ algorithm developed in this paper is given below:

\begin{algorithm}[H]
	\SetKwData{Left}{left}\SetKwData{This}{this}\SetKwData{Up}{up}
	\SetKwFunction{Union}{Union}\SetKwFunction{FindCompress}{FindCompress}
	\SetKwInOut{Input}{input}\SetKwInOut{Output}{output}\SetKwInOut{Tuning}{tuning parameter}\SetKwInOut{Function}{function} 
	\Input{Data $e_{1:T}$}
	\Output{$(\theta^i,s^i_{1:t})_{i=1:N}\sim p(\theta,s_{1:t}\mid e_{1:t})$, for $t=1,\dots,T$}
	\Tuning{number of particles $N$, particle filters $(pf_i)_{i=1:N}$}
    \Function{SMC$^2$ sampler $smc^2(e_{1:T})$}
    \BlankLine
    
	\tcp{Initialization}
    \For{$n\leftarrow 1$ \KwTo $N$}{
        Draw $\theta_n\sim p(\theta)$.
		Run particle filter $pf_n$ to obtain
		$s^n_{1:t_0}\sim \hat{p}_{\theta_n}(s^n_{1:t_0}\mid e_{1:t_0})$.
	}

	\tcp{Sequential update}
	\For{$t\leftarrow t_0+1$ \KwTo $T$}{
	
    \For{$n\leftarrow 1$ \KwTo $N$}{
        Propagate particle filter $pf_n$ and compute
        $\hat{p}_{\theta^n}(e_t\mid e_{1:t-1})$.
    }

    Normalize weights
    $\tilde{\alpha}^n_t\propto \hat{p}_{\theta^n}(e_t\mid e_{1:t-1})$.
    
    Compute
    \[
    \hat{L}_t=\sum_{n=1}^{N}\tilde{\alpha}^n_t
    \hat{p}_{\theta^n}(e_t\mid e_{1:t-1}).
    \]

    \uIf{Resampling required}{
    
        \For{$n\leftarrow 1$ \KwTo $N$}{
            Draw $k$ with $P(k=i)\propto \tilde{\alpha}^i_t$.
    
            Propose
            $(\theta^\star,s^\star_{1:t})
            \sim
            K_{pgibbs}(e_{1:t}, s^k_{1:t}, \theta^k)$.
    
            Set
            $(\theta^n,s^n_{1:t})
            :=
            (\theta^\star,s^\star_{1:t})$.
        }
    }
	}
	
    \KwRet $(\theta^i,s^i_{1:t})_{i=1:N}$ for $t=1,\dots,T$.
    
	\caption{Sequential Monte Carlo Squared (SMC$^2$)}
	\label{alg:SMC2}
\end{algorithm}

\subsection{Prediction} \label{subsec:cov_Prediction}

A key feature of the \acrshort{smc2} framework is that model parameters are updated sequentially over time. This enables prediction for future cases and deaths at each time step, which is particularly useful for model comparison; see Section \ref{subsec:cov_Modelcomparison}. In the \acrshort{smc2} context, prediction is performed naturally within the \acrlong{pf}. For particle $n$, given the state trajectory $s^n_{1:t}$, parameter $\theta$, and \acrshort{ode} state $O_t$, the next state $s^n_{t+1}$ is sampled from the transition dynamics. The \acrshort{ode} in Equation \eqref{eq:cov_Chp1_ode_SEEIIR} is then solved forward from $O_t$ using the transmission rate $\beta_{t+1}$ implied by $s^n_{t+1}$ and parameter $\theta$. The resulting quantities yield model-implied cases and deaths at time $t+1$, as defined in Section \ref{sec:cov_Model}.

\subsection{Model comparison} \label{subsec:cov_Modelcomparison}

In Section \ref{sec:cov_Application}, model comparison is performed to assess alternative observation models and to determine the number of latent regimes supported by the data. We consider both batch and sequential approaches. In the batch estimation case, we use the \acrfull{dic}, introduced in \cite{Spiegelhalter02} and the \acrfull{waic}, introduced in \cite{Watanabe10}.
Both criteria estimate the number of parameters in each model and provide a model score; with lower values indicating better fit. Both criteria can be computed with the parameter samples from a \acrshort{mcmc} run. The \acrshort{dic} is computed using:
\begin{equation}
    \operatorname{DIC}  = -2 log~p(e_{1:T} \mid \hat{\theta}_{Bayes}) + 2 \operatorname{p_{DIC}},
\label{eq:cov_BI_DIC}
\end{equation}
where $\hat{\theta}_{Bayes} = \mathbb{E}(\theta \mid e_{1:T})$ is the posterior mean given data $e_{1:T}$. For $\operatorname{p_{DIC}}$ we use half the variance of all likelihood estimates as suggested in the original \cite{Spiegelhalter02} manuscript, see also \cite{Gelman13}. 
For the \acrshort{waic} calculation, we follow \cite{Vehtari16}:
\begin{equation}
    \operatorname{WAIC} = -2 \operatorname{lppd} + 2 \operatorname{p_{WAIC}}, 
\label{eq:cov_BI_WAIC}
\end{equation}
where the \acrfull{lppd} can be interpreted as a training error and is computed as $\operatorname{lppd}= \sum_t^T log \left( \frac{1}{N} \sum_n^N p( e_t \mid \theta_n) \right)$ where T is the number of data points and N the number of \acrshort{mcmc} samples. The $\operatorname{p_{WAIC}}$ is computed as $\operatorname{p_{WAIC}} = \sum_t^T \operatorname{Var_{t, post}} \left( log~p(e_t \mid \theta ) \right)$, where $\operatorname{Var_{t, post}} \left( log~p(e_t \mid \theta ) \right)$ is the variance of $p(e_t \mid \theta )$ at time $t$ accross all $\theta$ samples.
Both $\operatorname{p_{DIC}}$ and $\operatorname{p_{WAIC}}$ serve as an appropriate penalty term based on the estimated model complexity within a score that approximates a model's predictive ability. In a predictive sequential model comparison context, it turns out that when using \acrshort{smc2} several quantities of interest arise as by-products of the estimation procedure, thus facilitating predictive model determination.

A desirable but typically computationally intractable quantity is the model's marginal likelihood $p(e_{1:T} ) = \int p(e_{1:T}, \theta)~d \theta$. Using \acrshort{smc2} an estimate for $p(e_t \mid e_{1:t-1})$ can be computed at each time step at practically no extra cost.
We denote $p(e_{t+1} \mid e_{1:t})$ as (one step ahead) \acrfull{pl} at $t+1$, $PL_{t+1}$, see \cite{Kastner16}, and note that it is straightforward to compute $\hat{p}(e_{1:t}) = \prod_{i=1}^t \hat{PL}_i$ after the \acrshort{smc2} run has finished.
However, direct comparison using these estimates is complicated by differences in the observation models across specifications.
An alternative approach to compute the $PL$ terms is based on the prediction step, which is described in Section \ref{subsec:cov_Prediction}. 
For \acrshort{hsmm}s, the posterior predictive distribution is defined as
$$
p(e_{t+1} \mid e_{1:t}) = \int p_{\theta}(e_{t+1} \mid s_{t+1}, s_{1:t}, e_{1:t}) ~ p_{\theta}(s_{t+1} \mid s_{1:t}, e_{1:t}) ~ p(s_{1:t}, \theta \mid e_{1:t})  ~ d s_{t+1}, s_{1:t}, \theta,
$$
where $ p_{\theta}(e_{t+1} \mid s_{t+1}, s_{1:t}, e_{1:t})$ can be evaluated and sampled from leading to a Monte Carlo estimate for the \acrlong{pl}, 
$\hat{PL}_{t+1} = \hat{p}(e_{t+1} \mid e_{1:t}) \approx \frac{1}{N} \sum_{n=1}^N p_{\theta_n}(e_{t+1} \mid e_{1:t}, s^n_{1:t+1})$,
where N is the number of samples. 
This estimate requires evaluation of the observation density and allows comparison based on components shared across models.

Direct model comparison can be conducted by computing the so-called \acrfull{clpbf}, which is derived from the cumulative sums of log \acrshort{pl}s. Given model $A$ and $B$, for $u > 0$, \acrfull{clpbf} reads
\begin{equation}
    \acrshort{clpbf}_{t+1:t+u} = log \left[ \frac{p_{A}(e_{t+u} \mid e_{1:t})}{p_{B}(e_{t+u} \mid e_{1:t})} \right] = \sum_{i=t+1}^{t+u} \left[\log  PL_{i}(A) - \log PL_{i}(B) \right],  
\end{equation}
where a positive value for \acrshort{clpbf} indicates evidence in favour of model A. In this context, if $t=0$ and $u=T$, the \acrshort{clpbf} reduces to the log \acrfull{bf}.
Note that when fitting real data to the \acrshort{smc2} algorithm in Section \ref{sec:cov_Application}, we usually use $t_0 > 1$ data points as training period to initialize the jitter kernels in a reasonable parameter region. 
This is especially useful in our case study as the reporting standard for the COVID-19 data at the beginning of the time series has been extremely noisy. The resulting, slightly amended, estimate for the comparison criteria is based on $\hat{p}(e_{t_{0}+1:T} \mid e_{1:t_0})= \prod_{t=t_0+1}^T \hat{PL}_{t}$; see \cite{Fong20} for some relevant discussion. 

\subsection{Simulation-based validation}

We conducted a simulation experiment to assess whether the proposed inferential procedure can recover model parameters under data generated from the assumed model class. A synthetic dataset of length 500 was generated from the preferred specification identified in the application section, using parameter values fixed at the corresponding posterior means. Posterior estimates were centred close to the generating values, with satisfactory mixing diagnostics across chains. These results provide supporting evidence that, under correctly specified model conditions, the proposed particle-based inference framework can recover the underlying parameters in practice. Full numerical summaries and diagnostic plots are reported in Supplementary Section 2.

\section{Application} \label{sec:cov_Application}

In this section, we apply the proposed methodology to the SARS-CoV-2 epidemic in the UK. In particular, we analyze the number of regimes needed to accurately capture the outbreak based on the model defined in Section \ref{sec:cov_Model}. Moreover, we consider whether expanding the evidence base yields a predictive advantage. The data for the proposed observation model includes fatalities, defined as the reported fatalities due to COVID-19 as interpreted by the UK government, as well as cases, which we take as the reported cases in the UK, see \cite{Coviddata22}. 
Observation model specifications often include either cases or fatalities. Cases can be noisy, are often a subset of the true burden and include lagging reporting periods. In contrast, fatalities are less noisy but typically necessitate additional complexity in the observation model, linking to the underlying transmission process. We combine both sources and test if this yields an advantage in terms of predictive performance against models that use fatalities only.

\subsection{Data, model dynamics, and prior specification} \label{subsec:cov_Data}

All data on reported fatalities, cases, and vaccinations in the UK were obtained from a publicly available website from the start of the recording period; see \cite{Coviddata22}. The initial data point for the estimation was taken at the first time index corresponding to at least ten reported fatalities; see Figure S3 of the Supplementary Material. Vaccination data have been set to zero where unavailable. The time horizon of the study was approximately 600 days. This period includes several waves of COVID-19 spikes in terms of reported cases. Moreover, estimates of \acrlong{ifr} values are available, see \cite{Brazeau22} for more details, and also Supplementary Figure S4.

The model consisting of both fatalities and cases in the observation distribution has parameters: $\theta = \{ \beta, \gamma, \epsilon, p, p_{thirdstate}, r, \psi, \phi_{cases}, \phi_{deaths} \}$ where $\beta, \gamma, \epsilon$ are \acrshort{ode} parameters in Equation \eqref{eq:cov_Chp1_ode_SEEIIR}. The vaccine efficacy parameter $\rho$ is fixed at $0.5$, while the delay parameter is set to $U=45$. These values were chosen to be broadly consistent with external evidence for the study period, reflecting higher efficacy against earlier SARS-CoV-2 variants and lower protection for later variants, together with a lag between first vaccination and effective immunity. We denote by $p$ and $p_{thirdstate}$ the transition distribution parameter and the same parameter for a non-recurring initial state respectively. This separate state captures the additional uncertainty at the start of the reporting period and significantly improves inference for all other parameter going forward. With $r$ and $\psi$ we denote the parameters of the Negative Binomial distribution used for the duration while $\phi_{cases}$ and $\phi_{deaths}$ absorb noise as described in more detail in Section \ref{sec:cov_Model}. 

The data
$e_t = \{ c^r_t, d^r_t \} \sim \operatorname{Neg. \hspace{0.1cm} Bin.}_{Alt} \left(c^*_t,c^*_t + \frac{c^{*2}_{t}}{\phi_c}\right) \times \operatorname{Neg. \hspace{0.1cm} Bin.}_{Alt} \left(d^i_t,d^i_t + \frac{{d^i}^{2}_{t}}{\phi_d}\right)$ depend on the model-implied fatalities, which are a function of the model-implied cases, which in turn are obtained from the \acrshort{ode} in Equation \eqref{eq:cov_Chp1_ode_SEEIIR}. This \acrshort{ode} depends on the latent state trajectory up to time $t$, and each latent state follows a \acrshort{hsmm}.
The \acrshort{ifr} values used to compute the model-implied fatalities defined in Section \ref{sec:cov_Model} are set to $\{ 0.01035, 0.0095, 0.007245, 0.004, 0.002 \}$ and the change point dates have been set to 2020-07-18, 2020-10-01, 2021-01-30 and 2021-06-01. Our choices are based on the results of \cite{Chatzilena22} and have been adjusted for the limited memory of the model-implied death computations. The same holds for the under-reporting score that is used in the observation model for the cases; see Figure S4 of the Supplementary Material.

The $\beta$ parameters are modelled on the log scale and form a vector of length equal to the number of latent states. We assign a multivariate normal prior subject to the ordering constraint $\beta_1 < \beta_2 < \cdots < \beta_K$ to ensure identifiability. For the four-state model, the prior mean vector is $\{\log(0.15), \log(0.4), \log(0.6), \log(1.2)\}$ with identity covariance matrix.
The elements of $\gamma$ have Gamma  $\operatorname{Gamma}(1600, \frac{1}{4000})$ and $\operatorname{Gamma}(2500, \frac{1}{5000})$ priors for shape and scale respectively. Similarly, Gamma priors have been chosen for $\epsilon \sim \operatorname{Gamma}(1000, \frac{1}{10000})$, $\phi_{cases} \sim \operatorname{Gamma}(2500, \frac{1}{500})$ and $\phi_{deaths} \sim \operatorname{Gamma}(2500, \frac{1}{500})$. Transition probability parameters $p$ and $p_{thirdstate}$ are assigned symmetric Dirichlet priors centred on equal weights, namely $p \sim \mathrm{Dirichlet}(k,\ldots,k)$, where $k$ is the number of latent states. The duration parameter satisfies $\psi \sim \mathrm{Beta}(0.5,0.5)$ for each state. For the Negative Binomial duration parameter $r$, Gamma priors are used with shape parameters $\{40,30,20\}$ in the four-state specification for recurring states and $28$ for the non-recurring state. Preliminary analyses indicated limited support for models with three or fewer regimes, both in terms of inferential stability and predictive performance. Hence, we focus on specifications with at least four latent states, and assess for potential additional regimes using the criteria in Section \ref{subsec:cov_Modelcomparison}.

\subsection{Results} \label{subsec:cov_Results}

\begin{figure}[htpb]
    \centering
	\includegraphics[ width=0.93\textwidth ]{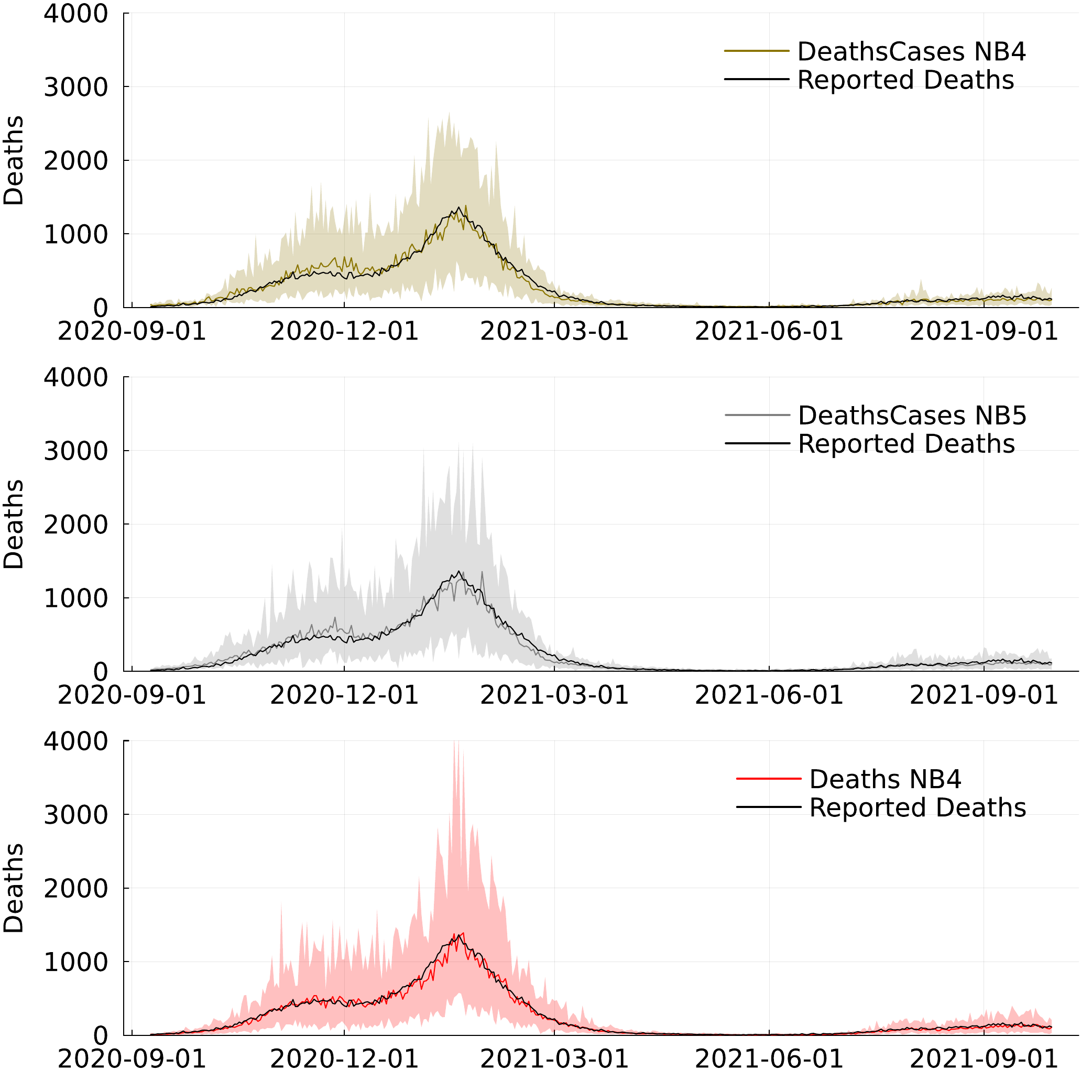}
	\caption{Daily predictions for the models described in Section \ref{sec:cov_Application}, with corresponding 95\% PIs, shown against the observed data. NB4 denotes a model with a Negative Binomial duration distribution and four regimes.}
    \label{fig:cov_Chp5_PredictionDailyComparisonSMC_1plot}
\end{figure}

We begin by comparing the models considered. A summary of all model comparison criteria is provided in Table \ref{table:cov_Chp5_Comparisons}. The lower two panels report the final cumulative log \acrshort{pl} values from the \acrshort{smc2} runs. These again show a clear advantage of the joint models (based fatalities and cases) over the fatalities-only benchmark, whose daily and weekly scores are $-26394$ and $-1985$, compared with values around $-16500$ and $-1844$ for the joint specifications. Figure S5 provides more information, provided the SMC$^2$ algorithm, on how the differences among the models evolved over time. Figure \ref{fig:cov_Chp5_PredictionDailyComparisonSMC_1plot}, which reports one-week-ahead forecasts of fatalities together with 95\% predictive intervals, provides further insight into the improvement offered by the joint models. Although point predictions are broadly similar across specifications, the predictive intervals are generally narrower for the models incorporating both fatalities and cases.

Between the two joint models, differences are small: the four-state model is slightly preferred for daily forecasts, whereas the five-state model has a negligible edge for weekly forecasts. For batch estimation, the \acrshort{dic} and \acrshort{waic} criteria, shown in the two upper panels, both favour the four-state fatalities--case model. In particular, its WAIC of $15705.3$ improves on the corresponding five-state value of $15837.2$, while the DIC values are $41011.8$ and $42093.1$, respectively.  Taking these results together, the four-state fatalities--cases specification offers the most favourable overall balance between fit, predictive performance, and parsimony, and is therefore used in the subsequent analysis.

\begin{table}[htpb]
\centering
  \begin{tabular}{rrrr} 
    \hline
   \multicolumn{4}{|c|}{\acrshort{dic} computations} \\
   \textbf{Model} & \textbf{DIC} & \textbf{ \acrshort{lppd}} & \textbf{$\mathbf{p_{DIC}}$} \\\hline
    Fatalities and Cases - 4 states & 41011.8 & -20043.5 & 462.4 \\
    Fatalities and Cases - 5 states & 42093.1 & -19272.0 & 1774.5 \\
  \end{tabular}

\centering
  \begin{tabular}{rrrr}
    \hline
    \multicolumn{4}{|c|}{\acrshort{waic} computations} \\
    \textbf{Model} & \textbf{WAIC} & \textbf{ \acrshort{lppd}} & \textbf{$\mathbf{p_{WAIC}}$} \\\hline
    Fatalities and Cases - 4 states & 15705.3 & -7768.9 & 83.7 \\
    Fatalities and Cases - 5 states & 15837.2 & -7749.2 & 169.4 \\
  \end{tabular}

\centering
  \begin{tabular}{rr}
    \hline
    \multicolumn{2}{|c|}{Daily sequential model choice} \\
    \textbf{Model} & \textbf{Daily cumulative log \acrshort{pl}} \\\hline
    Fatalities - 4 states & -26394 \\
    Fatalities and Cases - 4 states & -16523 \\
    Fatalities and Cases - 5 states & -16601 \\
  \end{tabular}

\centering
  \begin{tabular}{rr} 
    \hline
    \multicolumn{2}{|c|}{Weekly sequential model choice} \\
    \textbf{Model} & \textbf{Weekly cumulative log \acrshort{pl}} \\\hline
    Fatalities - 4 states & -1985\\
    Fatalities and Cases - 4 states & -1845 \\
    Fatalities and Cases - 5 states & -1843 \\
    \hline
  \end{tabular}

  \caption{Model comparison criteria for the specifications estimated on the real data in Section \ref{sec:cov_Application}. The two upper panels report batch model selection criteria, where lower values indicate better performance. The two lower panels report sequential model selection criteria, where higher values indicate better performance.}
  \label{table:cov_Chp5_Comparisons}
\end{table}

%
    
\begin{figure}[htp]
    \centering
	\includegraphics[ width=0.93\textwidth ]{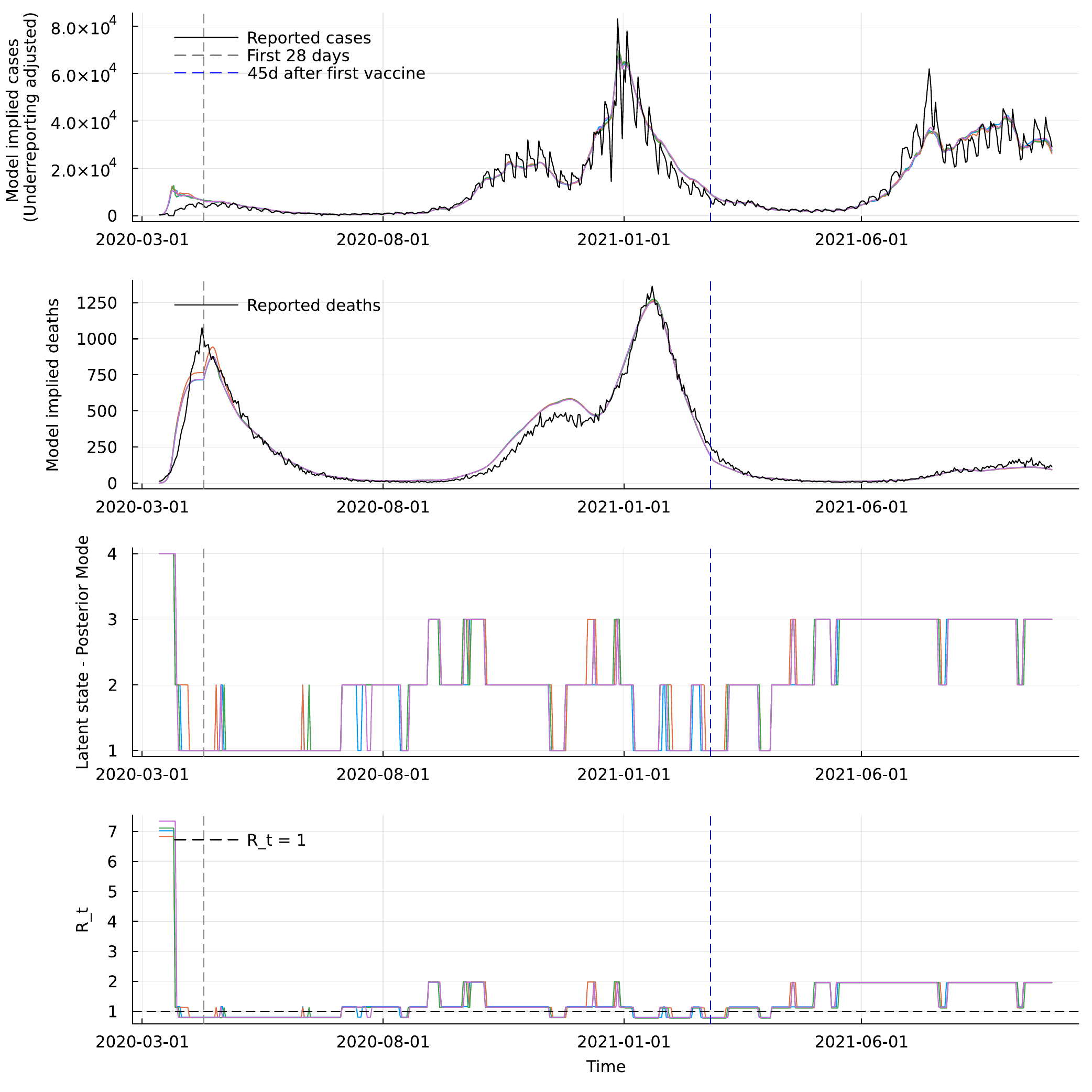}
	\caption{\acrshort{pmcmc} model implications for the preferred specification fitted to the real data in Section \ref{sec:cov_Application}. All computations use information available up to the corresponding time point on the x-axis. From top to bottom, the panels show: reported and model-implied cases (adjusted for under-reporting), reported and model-implied fatalities, the posterior mode of the latent regime, and the corresponding \acrshort{rt} trajectory. The grey vertical line marks the first 28 days, for which the full history required for model-implied fatalities is unavailable (see Section \ref{sec:cov_Model}); the blue vertical line marks the introduction of vaccination data into the \acrshort{ode} model.}
    \label{fig:cov_Chp5_ModelImplicationsPMCMC}
\end{figure}

Posterior summaries and output diagnostics for all parameters are reported in Table \ref{table:cov_Chp5_HSMM_PMCMC}. Summary plots comparing model-implied and reported cases and fatalities are shown in Figure \ref{fig:cov_Chp5_ModelImplicationsPMCMC}, which presents COVID-19 diagnostics using data available up to the current time. In the two upper panels, model-implied fatalities and cases, after accounting for under-reporting, track the reported data closely. The third panel shows the posterior mode of the latent state at each time point. Each latent state corresponds to a distinct transmission regime characterised by different transmission intensity and persistence. States are naturally ordered by posterior transmission intensity, with state 1 corresponding to the lowest transmission regime and state 4 to the highest.

\begin{table}[htp]
\centering
  \begin{tabular}{rrrrrrrrrrrr}
    \hline\hline
     $\mathbf{\theta}$ & \textbf{Mean} & \textbf{MCSE} & \textbf{SD} & \textbf{Rhat} & \textbf{Q2.5} & \textbf{Q25.0} & \textbf{Q50.0} & \textbf{Q75.0} & \textbf{Q97.5} \\\hline
    log $\beta_1$ & -1.72 & 0.0 & 0.08 & 1.01 & -1.89 & -1.78 & -1.72 & -1.67 & -1.57 \\
    log $\beta_2$ & -1.36 & 0.01 & 0.08 & 1.03 & -1.5 & -1.41 & -1.36 & -1.31 & -1.2 \\
    log $\beta_3$ & -0.81 & 0.0 & 0.09 & 1.0 & -0.99 & -0.87 & -0.81 & -0.76 & -0.63 \\
    log $\beta_4$ & 0.45 & 0.01 & 0.22 & 1.01 & 0.02 & 0.3 & 0.44 & 0.59 & 0.9 \\

    $\gamma_1$ & 0.45 & 0.0 & 0.04 & 1.0 & 0.37 & 0.42 & 0.45 & 0.48 & 0.54 \\
    $\gamma_2$ & 0.46 & 0.0 & 0.04 & 1.0 & 0.38 & 0.43 & 0.45 & 0.48 & 0.53 \\
  
    $\epsilon$ & 0.94 & 0.0 & 0.1 & 1.0 & 0.76 & 0.87 & 0.94 & 1.0 & 1.13 \\

    $p_1$ & 0.87 & 0.0 & 0.09 & 1.01 & 0.66 & 0.83 & 0.89 & 0.94 & 0.98 \\
    $p_2$ & 0.5 & 0.01 & 0.14 & 1.01 & 0.23 & 0.38 & 0.5 & 0.6 & 0.77 \\
    $p_3$ & 0.17 & 0.01 & 0.1 & 1.01 & 0.03 & 0.09 & 0.15 & 0.23 & 0.4 \\
  
    $p_{thirdstate, 1}$ & 0.35 & 0.01 & 0.15 & 1.0 & 0.09 & 0.23 & 0.34 & 0.45 & 0.67 \\
    $p_{thirdstate, 2}$ & 0.35 & 0.01 & 0.15 & 1.0 & 0.08 & 0.24 & 0.34 & 0.46 & 0.67 \\
  
    $r_1$ & 36.12 & 0.18 & 6.53 & 1.0 & 24.69 & 31.63 & 35.71 & 39.77 & 50.58 \\
    $r_2$ & 24.19 & 0.15 & 5.29 & 1.0 & 14.72 & 20.7 & 23.91 & 27.49 & 35.39 \\
    $r_3$ & 14.19 & 0.17 & 4.32 & 1.0 & 7.22 & 11.0 & 13.72 & 16.96 & 23.56 \\
    $r_4$ & 28.13 & 0.12 & 5.33 & 1.0 & 19.08 & 24.22 & 27.75 & 31.69 & 39.0 \\
  
    $\psi_1$ & 0.76 & 0.01 & 0.07 & 1.04 & 0.62 & 0.72 & 0.77 & 0.81 & 0.88 \\
    $\psi_2$ & 0.75 & 0.01 & 0.07 & 1.03 & 0.6 & 0.71 & 0.75 & 0.8 & 0.87 \\
    $\psi_3$ & 0.55 & 0.01 & 0.09 & 1.01 & 0.36 & 0.48 & 0.55 & 0.61 & 0.73 \\
    $\psi_4$ & 0.5 & 0.01 & 0.15 & 1.0 & 0.21 & 0.4 & 0.5 & 0.61 & 0.81 \\
  
    $\phi_{cases}$ & 4.91 & 0.0 & 0.11 & 1.0 & 4.68 & 4.83 & 4.91 & 4.99 & 5.13 \\
    $\phi_{deaths}$ & 5.25 & 0.0 & 0.12 & 1.0 & 5.02 & 5.17 & 5.26 & 5.33 & 5.49 \\
  
    \hline\hline
  \end{tabular}

\caption{Posterior summary statistics from four PMCMC chains for the \acrshort{hsmmem} with \acrshort{seir}-type \acrshort{ode} dynamics, four states, and a Negative Binomial duration distribution, estimated on the real data of Section \ref{sec:cov_Application}. Each chain was run for 1200 iterations with 700 burn-in iterations, yielding 2000 retained samples in total. Initial parameter values were sampled from the prior distributions.}
\label{table:cov_Chp5_HSMM_PMCMC}
\end{table}

Following the first date at which vaccination data are incorporated (blue dashed line) in the \acrshort{ode} model, a marked regime shift is observed. The posterior mode moves away from lower-transmission regimes (states one and two) towards regimes two and three, indicating a sustained increase in transmission intensity. The bottom panel shows the corresponding \acrshort{rt} values over time, which remain consistently higher after the vaccination effect is introduced. To reconcile these increased transmission levels with the contemporaneous vaccination effect, the model attributes part of the subsequent epidemic dynamics to factors consistent with increased transmissibility, such as the emergence of more transmissible SARS-CoV-2 variants.

Importantly, this structural change is identified sequentially, illustrating the practical value of the proposed online inference framework for real-time epidemic monitoring.

Finally, we note that if the Negative Binomial duration distribution were to approach the geometric case, the HSMM would reduce to a standard HMM. However, the estimated duration parameters depart materially from the geometric benchmark, indicating non-memoryless regime persistence and providing empirical support for the semi-Markov specification.

\section{Discussion} \label{sec:cov_Conclusion}

In this paper, we propose a new class of \acrshort{seir} models in which the transmission-rate parameter is piecewise constant and governed by a latent \acrshort{hsmm}. The resulting \acrshort{hsmm} formulation permits forecasting of both regime changes and regime durations, which may be useful for policy decision support. Moreover, the discrete latent structure yields an interpretable sequence of epidemic phases characterised by differing transmission intensity and persistence.

To estimate the latent \acrshort{hsmm}, we developed a particle-based inference framework tailored to this setting, combining \acrshort{pmcmc} and \acrshort{smc2} methods. The proposed computational schemes enabled inference using reported COVID-19 fatalities and infections in the UK. Our empirical findings indicate that (i) combining infections and fatalities improves predictive performance, (ii) using more than four latent regimes offers little additional benefit, and (iii) explicit duration modelling through the HSMM receives empirical support relative to the corresponding Markovian specification..

Several extensions are possible. Linking latent-state duration probabilities directly to observed covariates could allow inference on the timing and persistence of interventions such as lockdowns, conditional on the inferred regime structure. Additional data sources, when available, could also be incorporated into the observation model, including hospitalisations or intensive care admissions. Furthermore, the \acrshort{smc2} framework naturally provides model evidence estimates, enabling Bayesian model averaging across competing specifications with possible predictive gains. In the present work, however, we focused on individual models in order to preserve interpretability and transparent epidemiological communication. More broadly, this work contributes a flexible framework for epidemic dynamics together with a practical inference strategy for complex state-space models with semi-Markov structure and non-standard observations. These modelling and computational ideas may also support future investigations of optimal intervention policies, including decision-theoretic or reinforcement learning based approaches (e.g. \cite{Giacomo}).

\section{Software} \label{sec:cov_Sofware}


Replication code and data are available at \url{https://github.com/paschermayr/Publish_Covid19SSM}. Core Julia implementations are provided in Baytes.jl (\url{https://github.com/paschermayr/Baytes.jl}), with ODE computations based on DifferentialEquations.jl \citep{rackauckas17}.
\bibliographystyle{plainnat}
\bibliography{_references}

\end{document}